\documentclass[aapm,reprint,graphics]{revtex4-1}

\usepackage{graphicx}
\usepackage{amsmath}
\usepackage{array,booktabs}
\usepackage{color}

\usepackage{numcompress}

\begin{document}

\title{Quantifying allowable motion to achieve safe dose escalation in pancreatic SBRT}

\author{Yijun Ding, Ph.D.}
\author{Warren G.\ Campbell, Ph.D.}
\author{Moyed Miften, Ph.D.}
\author{Yevgeniy Vinogradskiy, Ph.D.}
\author{Karyn A.\ Goodman, M.D.}
\author{Tracey E. Schefter, M.D.}
\author{Bernard L.\ Jones, Ph.D.}
\email[e-mail correspondence: ]{bernard.jones@ucdenver.edu}
\affiliation{Department of Radiation Oncology, University of Colorado School of Medicine, Aurora, Colorado 80045}

\begin{abstract}
{\bf Background and purpose:} Tumor motion plays a key role in the safe delivery of Stereotactic Body Radiotherapy (SBRT) for pancreatic cancer. The purpose of this study was to use tumor motion data measured in patients to establish limits on motion magnitude for safe delivery of pancreatic SBRT, and to help guide motion-management decisions in potential dose-escalation scenarios.\\
{\bf Materials and methods:} Using 91 sets of pancreatic tumor motion data measured in patients, we calculated motion-convolved dose of the gross tumor volume (GTV), the duodenum and the stomach for 25 pancreatic cancer patients. We derived simple linear or quadratic models relating motion to changes in dose, and used these models to establish the maximum amount of motion allowable while satisfying error thresholds on key dose metrics. In the same way, we studied the effects of dose escalation and tumor volume on allowable motion. \\
{\bf Results:} In our patient cohort, the mean [minimum---maximum] allowable motion for 33/40/50 Gy to the PTV was 11.9 [6.3---22.4], 10.4 [5.2---19.1] and 9.0 [4.2---16.0] mm, respectively. Maximum allowable motion decreased as dose was escalated, and was smaller in patients with larger tumors. We found significant differences in allowable motion between different plans, suggesting a patient-specific approach to motion management is possible.\\
{\bf Conclusion:} The effects of motion on pancreatic SBRT are highly variable between patients and there is potential to allow more motion in certain patients, even in dose-escalated scenarios. In our dataset, a conservative limit of 6.3 mm would ensure safe treatment of all patients treated to 33 Gy in 5 fractions. \\

\textit{This manuscript was submitted to Practical Radiation Oncology}
\end{abstract}

\maketitle 

\section*{Introduction}

\indent At the time of diagnosis, approximately 50\% of pancreatic cancer patients present with distant metastases and 30\% of the patients have locally advanced disease [1]. Surgery is the preferred therapeutic option, but cannot be used in patients with either locally advanced or borderline resectable disease due to vascular involvement. Radiation plays a role in this setting, with the goal of either converting patients to surgical candidates or achieving local control of the tumor. Stereotactic Body Radiation Therapy (SBRT) has demonstrated promising early results [2, 3, 4]. However, the proximity of the pancreas to the radiosensitive organs of the abdomen (such as the small bowel, stomach, kidneys, and spinal cord) makes curative radiotherapy difficult.

\indent SBRT delivers large, ablative doses of radiation in only a few treatment fractions. SBRT can maximally spare the surrounding normal tissues by forming sharp dose gradients around the treatment volume [5, 6]. In recent years, multiple clinical trials have shown encouraging clinical outcomes of pancreatic cancer patients after SBRT [2, 7, 8, 3, 4, 9]. Furthermore, SBRT can potentially improve distant control by inducing immune response [10, 11, 12]. Current evidence suggests that dose escalation in SBRT may further improve patient outcomes [13, 14, 15, 16, 17], and many institutional trials of dose-escalated pancreatic SBRT are under way [18]. 

\indent However, dose escalation is limited by toxicities to the surrounding organs at risk (OARs). OARs in pancreatic SBRT include stomach and duodenum [19], because they are highly sensitive to radiation and adjacent to the pancreas. The location of tumor, duodenum and stomach in one patient are shown in Figure 1. Due to breathing, digestion, and heartbeat, the boundary between the tumor and nearby OARs is blurred. This internal target movement may lead to underdosage in parts of the tumor and overdosage to the OARs. As dose to the tumor escalates, these dose differences will grow larger, potentially increasing toxicity and decreasing local control. 

\begin{figure} [htb!]  
\centering
  \includegraphics[width=0.9\linewidth]{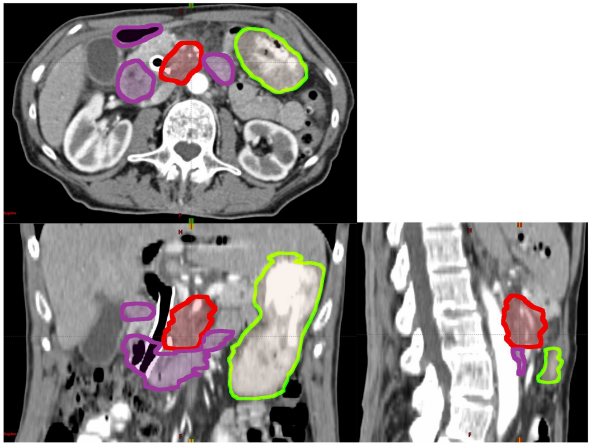}
\caption{The location of pancreatic tumor (GTV, red), duodenum (purple) and stomach (green) for a patient on a 3D CT scan.} \label{fig:CT_illustration}
\end{figure}

\indent To mitigate the detrimental effects on treatment due to the internal organ motion, techniques such as abdominal compression, respiratory gating, breath-hold, and real-time tumor tracking [20] can be applied. Each technique provides different tradeoffs between ease of use, patient comfort, and the magnitude of motion reduction. The AAPM task group 76 has advised an individual approach to respiratory management [20]; however, there is little clinical guidance on how to achieve this patient-specific management for pancreatic cancer patients. In this paper, we use tumor motion data measured in patients to a) quantify the effects of pancreatic motion on clinical dose distributions, and b) calculate the allowable tumor motion for individual patients. The purpose of our work is to develop a method for determining patient-specific motion limits. We utilize this method to answer three key clinical questions: How much motion is allowable during treatment while still achieving clinical dose constraints? Does allowable motion change accordingly in larger tumors or with dose escalation? Finally, is there a way to identify treatment plans that are more tolerant to motion, potentially allowing these patients to be treated with less onerous motion mitigation techniques?

\section*{Methods and Materials} \label{sec:methods}

\subsection*{Patients}

\indent Between November 2015 and July 2017, we treated 25 newly diagnosed pancreatic adenocarcinoma cases with Volumetric Modulated Arc Therapy (VMAT) to a dose of 33 Gy in five fractions using 6---10~MV photons. More details regarding these treatments can be found in a prior study [21]. Patients underwent two sets of CT scans prior to treatment: a free-breathing three-dimensional computed tomography (3D CT) scan for target definition and dose calculation and a four-dimensional CT (4DCT) scan to assess respiratory motion. Two motion-mitigation techniques were used during simulation and treatment. In 7 patients, abdominal compression was used during simulation and treatment to reduce the amplitude of motion. In the remaining 18 patients, respiratory gating was used. Gross tumor volumes (GTV) were drawn by the radiation oncologist on the planning CT (as shown in red in Figure~\ref{fig:CT_illustration}), and expanded to create a Planning Target Volume (PTV). The GTV-to-PTV expansion margins were typically 3-5 mm, and were determined by the oncologist based on local anatomy. PTV volumes were trimmed to avoid overlap with OARs. An Internal Target Volume (ITV) was not used; however, motion was taken into account in the design of the PTV. For this, the PTV contour was overlaid on the 4DCT, and compared against the observed range-of-motion of the tumor boundaries. For patients managed with gating, the range-of-motion included only the 30$\%$ to 70$\%$ phases of the 4DCT (end-exhale phases); whereas for patients managed with compression, this motion included all phases. If needed, the PTV was expanded further to encompass this range of motion. The EclipseTM Treatment Planning System (Varian Medical Systems, Palo Alto, CA) was used for planning and dose calculation. Data from these clinical cases were analyzed under an institutional review board (IRB) approved retrospective research protocol. 

\subsection*{Tumor trajectory}

\indent We utilized an existing pancreatic tumor motion database measured in patients to quantify the effects of motion on delivered dose. These tumor motion data were collected in a previous study of SBRT for pancreatic cancer [22]. This motion library included 91 datasets of daily pancreatic tumor motion derived from 19 patients. Raw CBCT projection images were utilized to calculate tumor motion data [23, 24, 25] with a previously developed technique. A summary of the technique is given as follows: 

\begin{enumerate}
\item In each projection image, the location of implanted intratumoral fiducial markers was determined with an automated template-matching algorithm \cite{campbell2017automated}.
\item A 3D Gaussian probability model of the tumor trajectory was formed with maximum-likelihood estimation.
\item For each projection, the 3D position of the tumor was estimated as the most likely position along the line of detection according to the 3D Gaussian model.
\item For each CBCT scan, a list of 892 tumor positions, which was also referred to as a tumor trajectory, was output in patient geometry. The coordinates of the tumor trajectories were in the anterior-posterior (AP), left-right (LR) and superior-inferior (SI) directions.
\end{enumerate}

\indent In each dataset, we calculated the time-resolved positions of the tumor throughout the CBCT, which were composed of 892 samples of 3D positions over 1 minute. Each sample was a 3D position estimated from one raw CBCT projection image (892 projections per CBCT, each acquired at 125~kVp, 1.5~mAs). These 19 patients were not included in the 25 treatment plans we studied in this paper.

\subsection*{Motion-convolved dose} 

\indent One of the main effects of organ motion in IMRT is, essentially, a blurring of the static dose distribution over the trajectory of the motion [26]. Methods based on convolution of the static dose distribution with a function representing the distribution of uncertainties from intratreatment organ motion have been proposed before [27, 28, 29]. In this paper, 3D tumor trajectories were used as the function representing internal organ motion. Motion-convolved dose was calculated under two assumptions: (1) the volume containing all structures of interest (the tumor, the duodenum and the stomach) was a rigid volume and there was no rotation of the organs; and (2) the static dose distribution was not affected by the internal organ motion.

\indent Dose volume histograms (DVHs) were calculated for GTV, duodenum and stomach. Note that we utilized coverage of the GTV (rather than the PTV) in our motion-convolved dose calculation, where the GTV is translated within the clinical dose distribution. This reflects the clinical scenario, where the tumor moves, and the ITV/PTV expansion is intended to encompass tumor motion and other uncertainties. The organ boundaries were generated from physician-drawn contours with shape-based interpolation [30]. For each patient and each tumor trajectory, the planned and motion-convolved dose distributions and DVHs were calculated and evaluated with several dose metrics. 

\subsection*{Relevant dose metrics}

\indent To determine the allowable motion for each patient, we computed the differences in dose metrics induced by motion. To date, pancreatic SBRT has been largely implemented at the single-institutional level, and as such there is some heterogeneity in terms of dose constraints and metrics reported in the literature [31, 32]. However, there is general agreement that the main constraints involve 1) coverage of the target volume and 2) maximum point dose and/or high dose constraints to the stomach and the duodenum. To understand the effects of motion on pancreatic SBRT, a range of dose metrics were selected to show the potential effects: (1) $D_{98}$, which is the dose covering $98\%$ of the GTV; (2) $D_{0.1cc\_d}$, which is the maximum dose covering at least 0.1~cm$^3$ of the duodenum; (3) $D_{0.1cc\_s}$, which is the maximum dose covering at least 0.1~cm$^3$ of the stomach; (4)$V_{33\_d}$, which is the volume of the duodenum that receives more than 33~Gy; and (5)$V_{33\_s}$, which is the volume of the stomach that receives more than 33~Gy.

\indent To get a broader understanding of the effect of motion on these plans, all 91-tumor trajectories were applied to each of the 25 treatment plans. In this analysis, the differences between the static and motion-convolved dose metrics ($\Delta$DMs, where each $\Delta{\text{DM}} = {\text{DM}}_{\text{motion}} - {\text{DM}}_{\text{static}}$) were calculated. In this paper, we refer to $\Delta$DMs as dosimetric errors. We divided the motion data into four bins (based on the magnitude of motion in SI direction) and calculated the mean and range of dosimetric errors for a group of patients. Prior work [22] has shown that the mean [minimum---maximum] target apparent motion ranges in SI direction for respiratory gating, abdominal compression and no motion mitigation are 5.5 [1.5---8.4], 8.5 [1.6---17.1] and 13.9 [4.7---35.5], respectively. The boundaries between the motion bins were at 4, 7, and 10 mm, such that the mean apparent motion ranges for respiratory gating, abdominal compression and no motion mitigation are in the second, third and fourth bins. 

\indent Personalized motion management requires knowledge about the effects of motion for each individual. To relate motion with dosimetric endpoints, we modeled the relation between $\Delta$DMs and the magnitude of motion in the SI direction for each patient. Based on the data collected from the 25 patients, we deduced a model, either linear or quadratic, for each dose metric. Data from each patient were fitted by the models to derive a set of patient-specific parameters. The models and the patient-specific parameters were used to predict $\Delta$DMs at any motion magnitude. Details of this method are given in the Supplemental Material. 

\subsection*{Allowable motion for individual patient}

\indent To help guide clinical interpretation of our results, we calculated limits for tumor motion based on a set of representative dosimetric criteria. We first chose a set of threshold values, which represented an excessive error in the dose metrics. The thresholds we chose are 2~Gy for $\Delta D_{98}$, 5~Gy for $\Delta D_{0.1cc\_d}$ and $\Delta D_{0.1cc\_s}$, 0.5~cm$^3$ for $\Delta V_{33\_d}$ and $\Delta V_{33\_s}$. These thresholds were used to calculate the upper limit on allowable motion ($L_{max}$) for each patient, which is the maximum magnitude of motion that satisfies all the above five error thresholds. The motion magnitude is represented by 95$\%$ confidence interval (CI) in the SI direction. 

\indent We explored the effects of dose escalation, tumor volume and varying dosimetric-error thresholds on allowable motion.

\textit{Dose escalation:} To study the allowable motion when the dose is escalated, we repeated the calculation for prescription doses of 40~Gy and 50~Gy. We assumed the planned dose distributions did not change and linearly scaled the original dose distribution up to the given value. To test the validity of this assumption, we re-planned 6 cases (the 3 most-sensitive and the 3 lease-sensitive to motion) to 50 Gy and calculated allowable motion using the re-planned dose distributions.

\textit{Tumor volume:} To investigate if there is correlation between the motion sensitivity of $\Delta$DMs and tumor volume $V_{GTV}$, we divided patients into two groups at the median of the 25 tumor volumes, which is 20~cm$^3$, and quantified the effects of motion separately for these two groups. These differences were assessed with an unpaired t-test. 

\textit{Sensitivity to dosimetric error thresholds:} In order to understand the effects of varying dosimetric-error thresholds on allowable motion, we performed a sensitivity analysis of the error thresholds on $L_{max}$. We calculated $L_{max}$ under varying dosimetric error thresholds. We fixed the thresholds for two dosimetric errors at our initial choices, which are $\max(\Delta D_{98})=2$~Gy, $\max(\Delta D_{0.1cc})=5$~Gy and $\max(\Delta V_{33})=0.5$~cm$^3$, and varied the third threshold, multiplying the initial choice by 0.5, 1 (no change) and 2.

\section*{Results} 
     
\begin{figure} [htb!]  
\centering
  \includegraphics[width=0.9\linewidth]{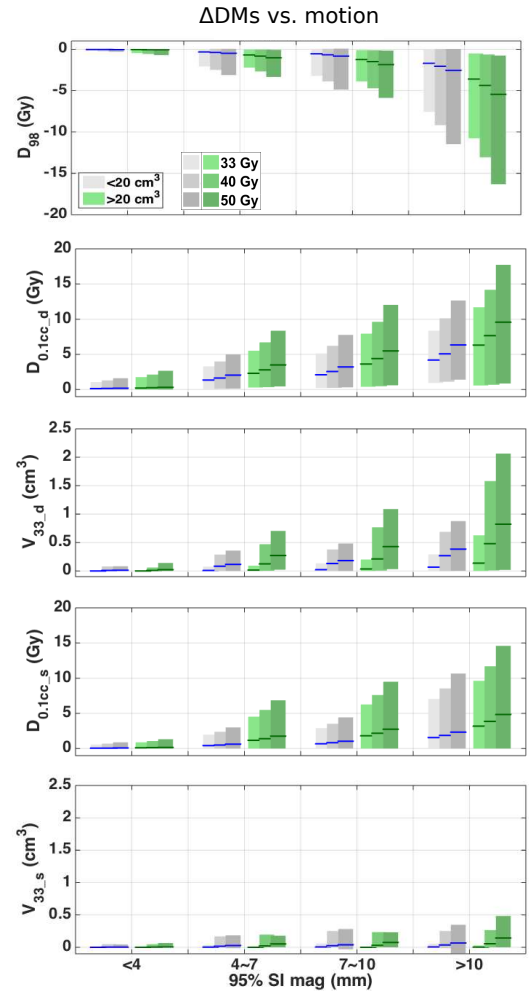}
\caption{$\Delta$DMs vs.\ motion for patients with small tumors (grey) and large tumors (green). From top to bottom, $\Delta D_{98}$, $\Delta D_{0.1cc\_d}$, $\Delta V_{33\_d}$, $\Delta D_{0.1cc\_s}$ and $\Delta V_{33\_s}$ vs.\ motion are plotted. Lines indicate mean; patches indicate $2.5^{th}$ to $97.5^{th}$ percentile. The data are displayed for three dose levels. The patients were divided into two groups based on tumor volumes, where the small tumor group ($V_{GTV}<20~cm^3$) has 13 patients and the large tumor group ($V_{GTV}\geq20~cm^3$) has 12 patients.}
\label{fig:DM_two_groups}
\end{figure}

\textit{Error in dose metrics ($\Delta$DMs):} The absolute values of the $\Delta$DMs increase with the magnitude of tumor motion and are larger for larger tumors, as shown in Figure~\ref{fig:DM_two_groups}, where data from all 25 patients are presented as box plots. The clinical dose distributions of patients with tumors larger than 20~cm$^3$ are affected more than those from smaller tumors by tumor motion. The tumor volume affected $\Delta V_{33\_d}$ more severely than error in other metrics considered in this paper. For reference, the average GTV volume was 20.0$\pm$10.7~cm3 (range 4.8---50.8) and the average PTV volume was 41.8$\pm$18.5~cm3 (range 17.4---94.6).

\begin{figure} [htb!]  
\centering
  \includegraphics[width=0.9\linewidth]{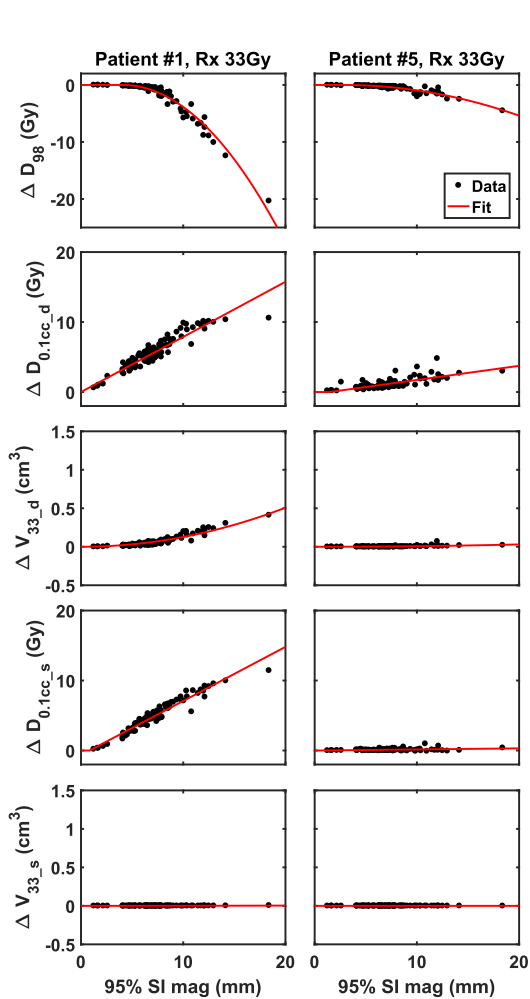}
\caption{$\Delta$DMs vs.\ magnitude of motion for individual patients. Data calculated from measured patient motion are plotted in black dots. The fitted curves are plotted in red line. }
\label{fig:fit_results}
\end{figure}

\indent The effects of motion on $\Delta$DMs are highly variable between patients. Figure~\ref{fig:fit_results} shows how $\Delta$DMs change as functions of SI motion for 2 representative patients. For each patient, the relations between $\Delta$DMs and SI motion were fitted by a set of models. Each patient had a set of fitting parameters. Details on the mathematical models and a quantitative measure of goodness of fit are discussed in Supplement~A. These personalized models, which relate dose metric errors to motion magnitude, provide foundations to calculate allowable motion for individual patient.

\indent In general, treatment plans broadly fall into two categories: those that are sensitive to tumor motion (eg.\ patient~$\#$1 as shown in the left column of Figure~\ref{fig:fit_results}) and those that were relatively stable as the magnitude of tumor motion increases (eg.\ patient~$\#$5 as shown in the right column of Figure~\ref{fig:fit_results}). 

\begin{figure} [htb!]  
\centering
  \includegraphics[width=\linewidth]{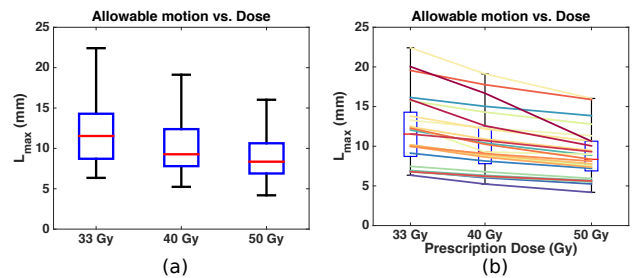}
\caption{Allowable motion vs. prescription dose for all 25 patients (a) and for each patient (b). (a): The boxplot shows the median (red center line), [$25^{th}$---$75^{th}$] percentile (blue box) and [minimum---maximum] range (black whiskers) of the 25 $L_{max}$ at each dose level. (b): Each colored-solid line represents one patient; the same boxplot in (a) is placed in the background for reference.}
\label{fig:mam_boxplot}
\end{figure}

\textit{Allowable motion ($L_{max}$):} The allowable motion, $L_{max}$, of all 25 patients are combined into a boxplot in Figure~\ref{fig:mam_boxplot} (a). The relation between $L_{max}$ and the prescription dose is presented for each patient in Figure~\ref{fig:mam_boxplot} (b), where each color represents one patient. The same boxplot as in Figure~\ref{fig:mam_boxplot} (a) is drawn in the background in Figure~\ref{fig:mam_boxplot} (b) for reference. The plots show that (1) different patients require different limits on target motion; (2) as prescription dose increases, allowable motion decreases; (3) inter-personal variation is larger than the variation caused by dose escalation.

\indent The allowable motion for patients with tumors larger than 20~cm$^3$ and patients with tumors smaller than 20~cm$^3$ are presented in Table~\ref{tb:mam_2groups_ttest}. As a side note, a $V_{GTV}$ of 20~cm$^3$ on average corresponds to $V_{PTV}$ of 42~cm$^3$. The mean [minimum---maximum] of $V_{GTV}$ for the two groups of patients is 28.3 [20.1---50.8] and 12.3 [4.8---19.4]~cm$^3$. Patients with larger tumors on average require a reduced limit on target motion.

\begin{table*}[tbh!]
\caption{Mean [minimum---maximum] allowable motion in mm and results of an unpaired two-sample student t-test to compare $L_{max}$ for patients with tumor volume larger than 20~cm$^3$ (12 patients) and smaller than 20~cm$^3$ (13 patients).}
\label{tb:mam_2groups_ttest}
\begin{center}
\begin{tabular}{ c c c c } 
\midrule
(mm) & 33~Gy  &  40~Gy  &  50~Gy   \\
\hline
\midrule
 $L_{max}$ (all patients) & 11.9 [6.3---22.4] & 10.4 [5.2---19.1] & 9.0 [4.2---16.0] \\
\midrule
$L_{max}$ ( $V>20~\text{cm}^3$)  & 10.0 [6.3---16.2] & 8.8 [5.2---15.0]  & 7.8 [4.2---13.9] \\ 
$L_{max}$ ( $V<20~\text{cm}^3$) & 13.7 [6.8---22.4]  & 12.0 [6.2---19.1] & 10.0 [5.6---16.0] \\
\hline
p-value & 0.03 & 0.03 & 0.07 \\ 
\hline
\end{tabular}
\end{center}
\end{table*}

     Figure 5 presents the mean and minimum—maximum range of the 25 Lmax for varying $\Delta$D98 (left), $\Delta$D0.1cc (middle) and V33 (right). As stated previously, the error threshold on one metric varies by multiplying the initial choice by 0.5, 1 (no change), and 2; the error thresholds on the remaining two metrics are fixed at our initial choices, which are max($\Delta$D98) = 2 Gy, max($\Delta$D0.1cc) = 5 Gy and max($\Delta$V33) = 0.5 cm$^3$.
     
\begin{figure*}  
\centering
\includegraphics[width=\linewidth]{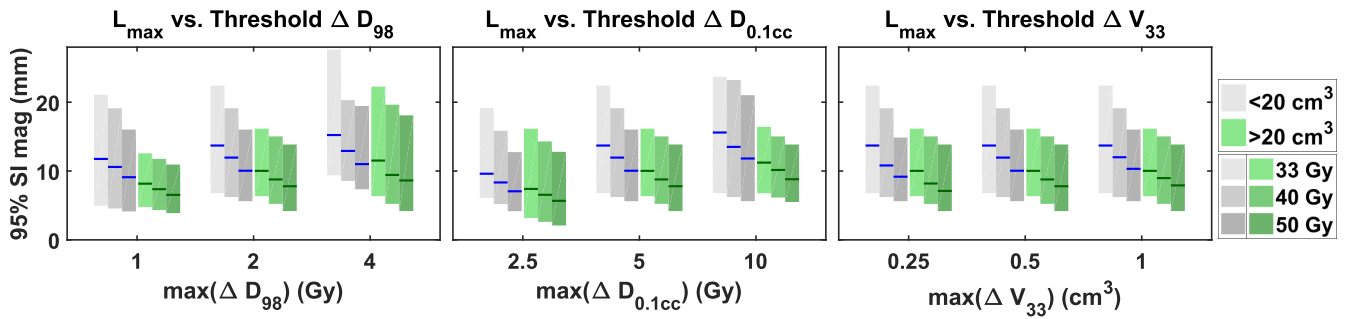}
  \caption{Upper limit on allowable motion ($L_{max}$) vs.\ varying error thresholds on $\Delta D_{98}$ (left), $\Delta D_{0.1cc}$ (middle) and $\Delta V_{33}$ (right). Each dosimetric-error threshold is varied at three values with the other two error thresholds fixed. When an error threshold is not varied, the value is fixed at:  $\Delta D_{98}=2$~Gy, $\Delta D_{0.1cc}=5$~Gy, and $\Delta V_{33}=0.5$~cm$^3$. Lines indicate mean of $L_{max}$; bars indicate range.}
  \label{fig:mam_dmc}
\end{figure*}

\textit{Sensitivity tests:} Figure~\ref{fig:mam_dmc} presents the mean and minimum---maximum range of the 25 $L_{max}$ for varying $\Delta D_{98}$ (left), $\Delta D_{0.1cc}$ (middle) and $V_{33}$ (right). As stated previously, the error threshold on one metric varies by multiplying the initial choice by 0.5, 1 (no change), and 2; the error thresholds on the remaining two metrics were fixed at our initial choices, which are $\max(\Delta D_{98})=2$~Gy, $\max(\Delta D_{0.1cc})=5$~Gy and $\max(\Delta V_{33})=0.5$~cm$^3$.

\indent In our sensitivity test of tissue deformation, we calculated the difference in the deformation vector field from the average deformation vector of the GTV. Differences in the deformation vector field averaged less than 1 mm at distances as high as 20~cm from the GTV.  The standard deviation of the deformation differences were less than 1~mm for duodenum, and less than 2~mm for the stomach.

\textit{Dose scaling:} For the 6 cases that we re-planned to test our assumption of linear dose scaling, the mean difference between allowable motion calculated from re-planned dose distributions and from scaled dose distributions ($L_{max\_scale} –L_{max\_replan}$) is $0.1\pm0.3$~mm, which on a percentage basis equates to $1.3\pm5.7\%$.

\section*{Discussion} \label{sec:discussion}

\indent As more institutions explore dose escalation in pancreatic SBRT, it is necessary to understand the potential risks and benefits. Motion is a first-order driver of uncertainty in abdominal SBRT, and the erratic nature of pancreatic motion [33, 34] makes accounting for this uncertainty difficult. A widely available technique to assess respiratory motion of pancreatic tumors is 4DCT; however, numerous studies have demonstrated that 4DCT underestimates the actual range of motion for pancreatic tumors [35, 36]. In this study, we developed a method to quantify the sensitivity of a given plan to motion, and tested this method using realistic tumor motion data measured in patients to quantify the effects of motion on our clinical dose distributions. By relating realistic motion data to dosimetric endpoints, we were able to establish limits on the amount of motion that can be allowed in this cohort without exceeding error thresholds. We found that the amount of allowable motion decreased as dose was escalated; however, this effect was small compared to inter-patients variations. We also found an inverse relationship between target volume and allowable motion. This makes intuitive sense, as in general, the boundaries of larger tumors are closer to the boundaries of OARs, and larger tumor volumes require more integral dose to provide adequate tumor coverage. The inverse relationship implies that the limits on motion should be tightened for large tumors (or relaxed for smaller tumors).

\indent In our dataset, a conservative limit of 6.3 mm motion ensured safe treatment of all patients at current dose levels. This value is highly dependent on the process through which the target volume was constructed, and the motion margins that were applied. While this specific result may apply only to our patient cohort, several results can be applied more broadly. First, we found considerable differences between patients (exceeding the differences due to dose escalation). This suggests that patient-specific dose escalation may be feasible, and that some patients could safely receive treatments that are more aggressive. Secondly, we have developed a method that would allow the determination of allowable motion prior to treatment. Taken together, one could envision a clinical scenario where, at the time of simulation, the patient undergoes comprehensive motion evaluation (using either fluoroscopy or CBCT in conjunction with a computational technique to extract motion data) [37, 24, 38]. Using the methods presented in this work, these motion data could then be used to evaluate the sensitivity of a given plan to motion, and an appropriate motion mitigation strategy could be implemented. In rare cases (with little predicted toxicity), it may even be possible to simply expand treatment margins to ensure coverage. On the other hand, in challenging dose-escalated scenarios, it may be appropriate to implement real-time imaging methods to increase the accuracy of treatment to better than 5 mm [39].

\indent Since unmitigated pancreatic motion is generally larger than 6.3 mm, our data imply that some form of motion mitigation is necessary. Minn et.\ al.\ [40] found that the range-of-motion in the SI direction can vary from 4.5 to 48.8~mm (mean 20.8~mm), and Feng et. al. [33] found a mean range-of-motion of $20\pm10$~mm. Clinically, motion management techniques, such as abdominal compression, respiratory gating, and real-time tumor-tracking [20], can be applied. In a prior study comparing gating and compression for pancreatic SBRT [22], it was found that abdominal compression resulted in an average of 8.5~mm of motion in the SI direction (range 2---17 mm), whereas the average apparent range of tumor motion during respiratory gating was 5.5 mm (range 2---8 mm). Taken together, these results suggest that respiratory gating may be the best one-size-fits-all approach for pancreatic SBRT. However, we also found significant differences between patients, and our data suggests that many patients could be treated safely under more relaxed conditions. 

\indent There are several important limitations to our work that must be considered. Our dose model assumes rigid motion within the abdomen, whereas previous work indicates that the pancreas can undergo deformation during respiration [33]. However, the deformation observed was up to 5 mm with a mean of 1---2 mm [33], which is small compared to the typical magnitude of pancreatic motion (1---2 cm). Furthermore, another study [41] compared rigid and deformable dose accumulation in liver SBRT and found that rigid dose accumulation led to dose errors averaging less than 1\% of prescription dose for tumor and normal tissue (ranging from -5.1\%---8.3\%). In our sensitivity analysis, this assumption of rigid motion resulted in average errors less than 1 mm.  Given these data, we argue that errors due to rigid assumptions are small compared to dosimetric changes caused by motion. However, future work on this topic should include the effects of deformation (perhaps using intrafraction images obtained from MR linacs). It is also important to understand the effects of inter-fraction anatomical changes. Day-to-day changes in patient anatomy (such as stomach/bowel gas) can cause dosimetric changes, and studies have found that adaptive re-planning in pancreatic SBRT can result in dosimetric benefits [42, 43]. The optimal approach may involve a combination of patient-specific motion management and adaptive replanning. As more institutions explore dose-escalated therapies, it is imperative to understand the benefits and risks of different techniques, and to choose the most appropriate approach.

\indent Another limitation is the ambiguity about which dose metrics are most relevant to pancreatic SBRT. Studies have shown that pancreatic SBRT toxicity is correlated with maximum dose received by an OAR ($D_{max}$), which refers to maximum dose to 1 cm$^3$ in one study [31] and maximum dose to 0.035 cm$^3$ in another [32]. We chose to use $D_{0.1cc}$ to represent $D_{max}$. Murphy et al.\ [31] has also shown that duodenum toxicity is correlated with the volume of duodenum receiving significant dose. Therefore, we chose to use $V_{33}$ to represent volume effect. The last metric we chose, which is $D_{98}$ of the GTV, represents tumor coverage. We anticipate that future clinical trials will provide more exact answers regarding pancreatic SBRT dose metrics; however, we chose a representative set of metrics that can be broadly applied to understand the relationship between motion, local control, and toxicity. There is also ambiguity about how much change in dose metrics is clinically relevant. We performed a sensitivity analysis (Figure~\ref{fig:mam_dmc}) to understand the impact of our choice of thresholds, but found that differences between patients ($\sim$1~cm) were much greater than differences due to choice of thresholds (2---3~mm). Once again, future clinical trials will provide better information in this regard, but our method provides a framework for relating these values to motion.

\indent It should be noted that, in this work, we simulated dose escalation by linearly scaling 33 Gy dose distributions to 40 or 50 Gy. This is somewhat unrealistic, as a clinical dose-escalated plan would be more optimized to spare normal tissue from harm. Better plan optimization would likely result in slightly more allowable motion. However, our data have shown a conservative limit of 4.2 mm of motion under linear-scaling to 50 Gy; and based on 6 re-planned cases, the difference in maximum allowable motion is $0.1\pm0.3$~mm under clinical dose escalation. The error introduced by the unrealistic dose-scaling is small compared to the difference between patients. Therefore, it is reasonable to claim that the differences in allowable motion between individual patients are much larger than the changes caused by dose escalation, as demonstrated by Figure~\ref{fig:mam_boxplot}.  Implementing a patient-specific motion mitigation approach requires knowing what makes a given plan more sensitive to motion. In this study, we found that allowable motion was significantly smaller in patients with larger tumors. Other features of the patients, such as the distance from the tumor to an OAR, may correlate with individual allowable motion as well. We investigated other potential factors (such as minimum distance from the tumor to duodenum or stomach), but were unable to find a clinically relevant way to predict the sensitivity of a given plan to motion. Future work will incorporate additional patient data, and attempt to develop metrics of motion sensitivity. There is potential for these sorts of metrics to be included in radiotherapy plan optimization, leading to the creation of plans that are more robust against tumor motion.

\section*{Conclusions} \label{sec:conclusion} 

\indent In a cohort of pancreatic cancer patients treated at our institution with 33~Gy/5~fractions, a conservative limit of 6.3~mm of motion prevents significant deviations in target coverage and OAR sparing from the radiotherapy plan. We observed tighter limits on allowable motion in patients with larger tumors, and for dose-escalated treatment plans. However, the largest variations in the effects of motion on $\Delta$DMs stem from differences between patients. Dose escalation can potentially be carried out for a sub-population of patients without additional clinical burden. We present a method to recognize this group of patients and provide guidance on personalized motion management.


\section*{Conflict of interest}
The University of Colorado School of Medicine and authors Campbell, Jones, and Miften have filed a provisional patent application for the fiducial marker tracking technique used in this work. 

\section*{Acknowledgment} 
This work was funded in part by the National Institute of Health under award number K12CA086913, the University of Colorado Cancer Center/ACS IRG $\#$57-001-53 from the American Cancer Society, the Boettcher Foundation, and Varian Medical Systems. These funding sources had no involvement in the study design; in the collection, analysis, and interpretation of data; in the writing of the manuscript; or in the decision to submit the manuscript for publication. 

\section*{References}
{\small \parindent0pt
[1] Hidalgo M. Pancreatic cancer. NEJM 2010;362(17):1605–17.

[2] Chuong MD, Springett GM, Freilich JM, Park CK, Weber JM, Mellon EA, et al. Stereotactic body radiation therapy for locally advanced and borderline resectable pancreatic cancer is effective and well tolerated. IJROBP 2013;86(3):516–22.

[3] Koong AC, Christofferson E, Le QT, Goodman KA, Ho A, Kuo T, et al. Phase II study to assess the efficacy of conventionally fractionated radiotherapy followed by a stereotactic radiosurgery boost in patients with locally advanced pancreatic cancer. IJROBP 2005;63(2):320–3.

[4] Koong AC, Le QT, Ho A, Fong B, Fisher G, Cho C, et al. Phase I study of stereotactic radiosurgery in patients with locally advanced pancreatic cancer. IJROBP 2004;58(4):1017–21.

[5] Rosati LM, Kumar R, Herman JM. Integration of stereotactic body radiation therapy into the multidisciplinary management of pancreatic cancer. Seminars in Radiation Oncology 2017;27(3):256 –67.

[6] Herman JM, Chang DT, Goodman KA, Dholakia AS, Raman SP, HackerPrietz A, et al. Phase 2 multi-institutional trial evaluating gemcitabine and stereotactic body radiotherapy for patients with locally advanced unresectable pancreatic adenocarcinoma. Cancer 2015;121(7):1128–37.

[7] Mahadevan A, Miksad R, Goldstein M, Sullivan R, Bullock A, Buchbinder E, et al. Induction gemcitabine and stereotactic body radiotherapy for locally advanced nonmetastatic pancreas cancer. IJROBP 2011;81(4):e615–22.

[8] Chang DT, Schellenberg D, Shen J, Kim J, Goodman KA, Fisher GA, et al. Stereotactic radiotherapy for unresectable adenocarcinoma of the pancreas. Cancer 2009;115(3):665–72.

[9] Ceha HM, van Tienhoven G, Gouma DJ, Veenhof CH, Schneider CJ, Rauws EA, et al. Feasibility and efficacy of high dose conformal radiotherapy for patients with locally advanced pancreatic carcinoma. Cancer 2000;89(11):2222–9.

[10] Twyman-Saint Victor C, Rech AJ, Maity A, Rengan R, Pauken KE, Stelekati E, et al. Radiation and dual checkpoint blockade activates nonredundant immune mechanisms in cancer. Nature 2015;520(7547):373.

[11] Golden EB, Chhabra A, Chachoua A, Adams S, Donach M, FentonKerimian M, et al. Local radiotherapy and granulocyte-macrophage colony-stimulating factor to generate abscopal responses in patients with metastatic solid tumours: a proof-of-principle trial. Lancet Oncology 2015;16(7):795–803.

[12] Lee Y, Auh SL, Wang Y, Burnette B, Wang Y, Meng Y, et al. Therapeutic effects of ablative radiation on local tumor require CD8+ T cells: changing strategies for cancer treatment. Blood 2009;114(3):589–95.

[13] Brown JM, Brenner DJ, Carlson DJ. Dose escalation, not new biology, can account for the efficacy of SBRT with NSCLC. IJROBP 2013;85(5):1159.

[14] Golden DW, Novak CJ, Minsky BD, Liauw SL. Radiation dose $\geq$54 Gy and CA 19–9 response are associated with improved survival for unresectable, non-metastatic pancreatic cancer treated with chemoradiation. Radiation Oncology 2012;7(1):156.

[15] Park HJ, Griffin RJ, Hui S, Levitt SH, Song CW. Radiation-induced vascular damage in tumors: implications of vascular damage in ablative hypofractionated radiotherapy (SBRT and SRS). Radiation research 2012;177(3):311–27.

[16] Stinauer MA, Kavanagh BD, Schefter TE, Gonzalez R, Flaig T, Lewis K, et al. Stereotactic body radiation therapy for melanoma and renal cell carcinoma: impact of single fraction equivalent dose on local control. Radiation Oncology 2011;6(1):34.

[17] Timmerman R, Paulus R, Galvin J, Michalski J, Straube W, Bradley J, et al. Stereotactic body radiation therapy for inoperable early stage lung cancer. JAMA 2010;303(11):1070–6.

[18] Available online at www.clinicaltrials.gov [Acccessed: 10/20/2017]; 2017. Searching terms: SBRT AND dose escalation AND pancreatic cancer.

[19] Hoyer M, Roed H, Sengelov L, Traberg A, Ohlhuis L, Pedersen J, et al. Phase-II study on stereotactic radiotherapy of locally advanced pancreatic carcinoma. Radiotherapy and oncology 2005;76(1):48–53.

[20] Keall PJ, Mageras GS, Balter JM, Emery RS, Forster KM, Jiang SB, et al. The management of respiratory motion in radiation oncology report of AAPM Task Group 76. Medical physics 2006;33(10):3874–900.

[21] Vinogradskiy Y, Goodman KA, Schefter T, Miften M, Jones BL. The clinical and dosimetric impact of real-time target tracking in pancreatic SBRT. IJROBP, 2019;103(1):268-275.

[22] Campbell WG, Jones BL, Schefter T, Goodman KA, Miften M. An evaluation of motion mitigation techniques for pancreatic SBRT. Radiotherapy and Oncology 2017;124(1):168-173.

[23] Campbell WG, Miften M, Jones BL. Automated target tracking in kilovoltage images using dynamic templates of fiducial marker clusters. Medical physics 2017;44(2):364–74.

[24] Jones BL, Westerly D, Miften M. Calculating tumor trajectory and dose-of-the-day using cone-beam CT projections. Medical physics 2015;42(2):694– 702.

[25] Poulsen PR, Cho B, Keall PJ. A method to estimate mean position, motion magnitude, motion correlation, and trajectory of a tumor from cone-beam CT projections for image-guided radiotherapy. IJROBP 2008;72(5):1587–96.

[26] Bortfeld T, Jokivarsi K, Goitein M, Kung J, Jiang SB. Effects of intrafraction motion on IMRT dose delivery: statistical analysis and simulation. PMB 2002;47(13):2203.

[27] Lujan AE, Larsen EW, Balter JM, Ten Haken RK. A method for incorporating organ motion due to breathing into 3D dose calculations. Medical physics 1999;26(5):715–20.

[28] McCarter S, Beckham W. Evaluation of the validity of a convolution method for incorporating tumour movement and set-up variations into the radiotherapy treatment planning system. PMB 2000;45(4):923.

[29] Lens E, Kotte AN, Patel A, Heerkens HD, Bal M, van Tienhoven G, et al. Probabilistic treatment planning for pancreatic cancer treatment: prospective incorporation of respiratory motion shows only limited dosimetric benefit. Acta Oncologica 2017;56(3):398–404.

[30] Herman GT, Zheng J, Bucholtz CA. Shape-based interpolation. IEEE Computer Graphics and Applications 1992;12(3):69–79.

[31] Murphy JD, Christman-Skieller C, Kim J, Dieterich S, Chang DT, Koong AC. A dosimetric model of duodenal toxicity after stereotactic body radiotherapy for pancreatic cancer. IJROBP 2010;78(5):1420–6.

[32] Goldsmith C, Price P, Cross T, Loughlin S, Cowley I, Plowman N. Dose-volume histogram analysis of stereotactic body radiotherapy treatment of pancreatic cancer: A focus on duodenal dose constraints. In: Seminars in radiation oncology; vol. 26. Elsevier; 2016, p. 149–56.

[33] Feng M, Balter JM, Normolle D, Adusumilli S, Cao Y, Chenevert TL, et al. Characterization of pancreatic tumor motion using cine MRI: surrogates for tumor position should be used with caution. IJROBP 2009;74(3):884–91.

[34] Heerkens HD, van Vulpen M, van den Berg CA, Tijssen RH, Crijns SP, Molenaar IQ, et al. MRI-based tumor motion characterization and gating schemes for radiation therapy of pancreatic cancer. Radiotherapy and Oncology 2014;111(2):252 –7.

[35] Lens E, van der Horst A, Kroon PS, van Hooft JE, Dávila Fajardo R, Fockens P, van Tienhoven G, Bel A. Differences in respiratory-induced pancreatic tumor motion between 4D treatment planning CT and daily cone beam CT, measured using intratumoral fiducials. Acta Oncologica. 2014;53(9):1257-64. 

[36] Rankine LJ, Wan H, Parikh PJ, Maughan NM, Poulsen PR, Santanam L. Cone Beam CT Measures Abdominal Tumor Motion Better Than 4DCT and Equal to On-table Fluoroscopy. IJROBP. 2015;93(3):S116. 

[37] Poulsen PR, Cho B, Keall PJ. Real-time prostate trajectory estimation with a single imager in arc radiotherapy: a simulation study.PMB 2009;54(13):4019. 

[38] Berbeco RI, Nishioka S, Shirato H, Chen GT, Jiang SB. Residual motion of lung tumours in gated radiotherapy with external respiratory surrogates. PMB 2005;50(16):3655. 

[39] Keall PJ, Nguyen DT, O’Brien R, Caillet V, Hewson E, Poulsen PR, et al. The first clinical implementation of real-time image-guided adaptive radiotherapy using a standard linear accelerator. Radiotherapy and Oncology 2018;127(1):6-11.

[40] Minn AY, Schellenberg D, Maxim P, Suh Y, McKenna S, Cox B, et al. Pancreatic tumor motion on a single planning 4D-CT does not correlate with intrafraction tumor motion during treatment. American journal of clinical oncology 2009;32(4):364–8.

[41] Velec M, Moseley JL, Eccles CL, Craig T, Sharpe MB, Dawson LA, et al. Effect of breathing motion on radiotherapy dose accumulation in the abdomen using deformable registration. IJROBP 2011;80(1):265–72.

[42] Bohoudi O, Bruynzeel A, Senan S, Cuijpers J, Slotman B, Lagerwaard F, et al. Fast and robust online adaptive planning in stereotactic MR-guided adaptive radiation therapy (SMART) for pancreatic cancer. Radiotherapy and Oncology 2017;125(3):439–44.

[43] Li Y, Hoisak JD, Li N, Jiang C, Tian Z, Gautier Q, et al. Dosimetric benefit of adaptive re-planning in pancreatic cancer stereotactic body radiotherapy. Medical Dosimetry 2015;40(4):318–24.

}

\section*{Supplement A}
The mathematical models are chosen based on two merits: (1) ability to account for the trend of the data, and (2) simplicity. We chose either linear or quadratic models that only require maximum of two fitting parameters. We present the models and the goodness-of-fit statistics. 

\subsection{$\Delta D_{98}$ }
\indent We used a piecewise-parabola function to describe the trend of $\Delta D_{98}$ with changing motion magnitude. The piecewise parabola function was given by

\begin{equation}
	\Delta D_{98} (x) = 
	\begin{cases}
		0\,,\qquad\qquad\qquad\,\,\,\,\,\,\, \text{if $x<p_2$} \,;\\
		p_1 \cdot (x-p_2)^2\,,\qquad \text{if $x\geq p_2$}\,;
	\end{cases}        
\end{equation}
where $x$ is the $95\%$ confidence interval (CI) of motion range in superior-inferior (SI) direction, $p_1$ and $p_2$ are two fitting parameters. 

\subsection{$\Delta D_{0.01cc}$ }
\indent We used a piecewise linear function to describe the trend of $\Delta D_{0.01cc}$. The piecewise linear function was given by

\begin{equation}
	\Delta D_{0.01cc}(x) = 
	\begin{cases}
		0\,,\qquad\qquad\qquad\,\,\,\,\,\,\, \text{if $x<p_2$} \,;\\
		p_1 \cdot (x-p_2)\,,\qquad \text{if $x\geq p_2$}\,;
	\end{cases}        
\end{equation}
where $x$ is the $95\%$ CI of motion range in SI direction, $p_1$ and $p_2$ are two fitting parameters. This model is used for $\Delta D_{0.01cc}$ in both duodenum and stomach.

\subsection{$\Delta V_{33}$ }
\indent The trend of $\Delta V_{33}$ was described by a parabola function of form

\begin{equation}
	\Delta V_{33}(x) = p_1 \cdot x^2,       
\end{equation}
where $x$ is the $95\%$ CI of motion range in SI direction, $p_1$ is the only fitting parameter. This model is used for $\Delta V_{33}$ in both duodenum and stomach.

\subsection{Goodness-of-fit statistics}

After fitting data with our models, we first visually inspected the goodness of fit. As shown in example plots of two patients, the models fitted the trend of the data qualitatively. To evaluate the goodness of fit quantitatively, we calculated the root mean squared error (RMSE). For each metric, we calculated the mean and standard deviation of the RMSE over 25 patients and the results are presented in Table~\ref{tb:fit_rmse}. The RMSE values were small compared to the range of the $\Delta$DMs, which indicates that the models were useful for prediction.

\begin{table}[h!]
\caption{The mean and standard deviation of RMSE over 25 patients. }
\label{tb:fit_rmse}
\begin{center}
\begin{tabular}{ c c c c c c } 
\midrule
 Rx dose (Gy) & 33  &  40  &  50   \\
\hline
$\Delta D_{98}$ (Gy)            & 0.3$\pm$0.2 & 0.4$\pm$0.3  & 0.5$\pm$0.3 \\ 
$\Delta D_{0.1cc\_D}$ (Gy)  & 0.7$\pm$0.2 & 0.9$\pm$0.3  & 1.1$\pm$0.3 \\ 
$\Delta V_{33\_D}$ (cm$^3$)& 0.0$\pm$0.0 & 0.1$\pm$0.0  & 0.1$\pm$0.1 \\ 
$\Delta D_{0.1cc\_S}$ (Gy)   & 0.4$\pm$0.3 & 0.5$\pm$0.3  & 0.6$\pm$0.4 \\
$\Delta V_{33\_S}$ (cm$^3$) & 0.0$\pm$0.0 & 0.0$\pm$0.0  & 0.0$\pm$0.0 \\
\hline
\end{tabular}
\end{center}
\end{table}

\end{document}